\documentstyle[sprocl]{article}

\def\Journal#1#2#3#4{{#1} {\bf #2}, #3 (#4)}

\def\MOD{{\em Mod. Phys. Lett} A}

\def\NPB{{\em Nucl. Phys.} B}
\def\PLB{{\em Phys. Lett.}  B}
\def\PRL{\em Phys. Rev. Lett.}
\def\PRD{{\em Phys. Rev.} D}

\newcommand{\beba}{\begin{equation}\begin{array}{lcl}}
\newcommand{\eaee}{\end{array}\end{equation}}
\newcommand{\bea}{\begin{eqnarray}}
\newcommand{\eea}{\end{eqnarray}}
\newcommand{\ba}{\begin{array}}
\def\nn{\nonumber}

\def\d{\delta}

\def\k{\kappa}
\def\l{\lambda}
\def\m{\mu}
\def\n{\nu}

\def\r{\rho}
\def\s{\sigma}

\def\t{\tau}

\def\be{\begin{equation}}
\def\ee{\end{equation}}
\def\bea{\begin{eqnarray}}
\def\eea{\end{eqnarray}}

\begin{document}

\title{Scales and  Cosmological Applications of M Theory}

\author{Karim Benakli}

\address{Phys. Dept.  Texas A \& M, College Station \\TX
77843, USA.\\E-mail: karim@chaos.tamu.edu}

\maketitle\abstracts{ I review recent results in three topics of the M-world:
(i) Scales. (ii) New dark matter candidates. 
(iii) Cosmological solutions from p-branes. The three topics are discussed 
in the framework of Ho\v{r}ava-Witten compactifications. Part (iii) 
includes comments on 
cosmological solutions in M-theory describing nucleation of universes through instanton effects and expansions toward asymptotically 
flat or  anti-de-Sitter spaces.}

\section{Introduction: }

Recent works have provided evidence for the existence of a fundamental 
theory (M-theory) unifying all known vacuua of string theory as well as 
eleven dimensional supergravity. This allows us to hope that a mathematical
structure unifying the algebraic structure of quantum mechanics and 
the geometrical approach of general relativity  in a consistent way 
exists. M-theory in the way it is formulated today assumes that the law of
quantum mechanics are valid at length scales far much smaller that those
probed experimentally. At very short scales particles cease to be point-like 
and complicate dynamics of extended objects might lead to modifications 
of the form of the fundamental laws. However, mathematical consistency 
will not replace checking experimental predictions of M-theory to prove its 
relevance to the description of our world. Away from such an ambitious  
and probably premature program, it is useful to study the 
different phenomenologies that arise from different vacuua of M-theory as 
they provide a new sight on problems of low energy physics (as the 
naturality of the structure and parameters of the standard model) and 
hopefully this will also have a feedback on mathematical physics.

Models with chiral $N=1$ supergravity effective theory can be obtained
by  compactification of M-theory  on orbifolds. The simplest example
is the Ho\v{r}ava-Witten \cite{HW} supergravity. In the infrared limit,
the geometry is that of an eleven-dimensional space with two
boundaries.  The low energy  effective theory in the bulk contains a
graviton supermultiplet formed by the degrees of freedom of 
graviton, gravitino and a three form $C_{IJK}$ whose four-form field strength 
we denote by
$G$. On each boundary there are nine-branes carrying an $E_8$
Yang-Mills supermutiplet. There are two scales appearing in this model: the
eleven-dimensional Planck mass $M_{11}$ and the size of the $S^1/Z_2$
segment.

In section 2 we discuss the scales which govern the phenomenology
arising from this theory in the case of compactification on
non-singular Calabi-Yau manifolds. Many  authors have explicitly 
discussed in the past the
expected sizes for the different  scales appearing in the
Ho\v{r}ava-Witten model\cite{witten,BD,KP,anton,nano1,CKM,nilles}.  However, 
they  focused on the standard embedding in which case   the
segment $S^1/Z_2$ has a maximal length. This critical value depends on
details of the compactification considered and  thus  was  never
computed explicitly.  I believe it is useful to  reconsider this issue in
a more generic Calabi-Yau context.  In section 3 we consider the possibility
that dark matter originates on the hidden wall or in the bulk of the theory. 
We outline the main properties and possible signature. A detailed study will 
appear elsewhere\cite{benf}. Section 4 is devoted to attempts to obtain 
simple cosmological solutions in M-theory. We first discuss a toy model in
the Ho\v{r}ava-Witten supergravity context obtained by rotating an euclidean 
instanton to Lorentzian signature. This model exhibits two 
interesting features: spontaneous compactification and difference in early 
time behavior of  the boundaries. We comment at the end on the possibility to 
generate solutions which describe asymptotically anti-de-Sitter spaces.

\section{Scales revisited}

One looks for a solution of the equations of motion of M-theory
describing an eleven dimensional manifold, with boundaries
separated by a segment of length\footnote{ The lengths are
measured using the eleven-dimensional metric.} $\pi \rho$, compactified on a
Calabi-Yau six-fold $CY$ of volume $V$. Following Witten\cite{witten}, this
solution can be organized as an expansion in the dimensionless
parameter $\rho M_{11}^{-3}/V^{2/3}$.  At the lowest order   our vacuum
has the geometrical interpretation of  ${\cal M}_4 \times CY \times
S^1/Z_2$. At higher orders in ${\rho  M_{11}^{-3}/V^{2/3}}$  the
factorization is lost and  the volume of the Calabi-Yau space becomes  a
function of  the coordinate parametrizing the $S^1/Z_2$ segment. More
precisely, within the  approximations of \cite{witten} the volumes of $CY$
seen by the observable sector\footnote{ We will use the subscripts $o$
for parameters of the observable sector and $h$ for those of the hidden sector.} $V_{o}$ and the one on the hidden sector $V_{h}$ 
are given by\cite{HW,BD}:  
\be  
V_{o} =
V \left(1 + \left(\frac {\pi}{2}\right)^{4/3} a_o {\rho 
M_{11}^{-3}\over V^{2/3}}\right) 
\label{volo}
\ee  and  \be  V_{h} = V \left(1 + \left( \frac {\pi}{2}\right)^{4/3} a_h
{\rho M_{11}^{-3}\over V^{2/3}} \right)
\label{volh}
\ee  thus 
\be  V_{h} = V_{o} + \left(\frac {\pi}{2}\right)^{4/3}
(a_h-a_o) M_{11}^{-3} V^{1/3} \rho 
\label{diffvol}
\ee  
where now $V$ is the (constant) lowest order value for the volume of the 
Calabi-Yau manifold and $a_{o,h}$ are model-dependent constants that 
we will discuss in some details below.

Due to non trivial gauge configuration (``instanton'' like) in the
internal Calabi-Yau manifold, the $E_8$ gauge  symmetries on the
boundaries are replaced by the gauge groups $G_o$ at $x^{11}=0$ and
$G_h$  at $x^{11}=\pi \rho$.  We would like to
consider the possibility of embbeding the minimal supersymmetric 
standard model   in this
M-theory framework. Under the assumption of a desert scenario 
LEP data \cite{pre,LEP} suggest unification at a scale 
$M_{GUT}\sim 2\cdot 10^{16}$ GeV. In an abuse of language, we will refer to 
the scale and group of unification as GUT scale and group while bearing in 
mind that there might be no new unifying gauge symmetry and $G_o$ could be 
just  
the standard
model group  $SU(3)\times SU(2)\times U(1)$. If $M_{GUT}>>V_{o}^{-1/6}$ then 
the theory becomes higher
dimensional before reaching $M_{GUT}$ and the four dimensional
field theoretical prediction of  unification is no more valid. In presence 
of twisted states arising from orbifold 
singularities, field theory is not reliable and one needs to use 
fully M-theoretical
computations. Typically  $M_{GUT}$ is expected to be roughly of the
order of the mass of the lightest heavy mode on the Calabi-Yau
manifold. If we suppose that  the standard model resides on the
$x_{11}=0$ wall, we have:  
\be  
M_{GUT}\sim  c  V_{o}^{-1/6} \; ,
\label{mgutc}
\ee where $c$ is a model dependent constant. It might involve
geometrical factors  as well as contribution from the gauge symmetry
breaking (Wilson lines...).  However explicit computations in the
weakly coupled heterotic string case seem to indicate that the natural value
remains $c\sim 1$ \cite{KP}. A similar discussions  holds for the scale
associated with the hidden group $G_h$ on the other side  of the
universe.

Three other quantities of phenomenological interest are the Newton
constant $G_N$ and the values of the coupling constants $\alpha_{o}$
of $G_o$ (that we wish to identify with the GUT group and its coupling
constant) and $\alpha_{h}$ of $G_h$. They are given by\cite{witten,LU}:

\be G_N  = {1\over 16 \pi^2}{1\over M_{11}^9 \rho \langle V\rangle} \,,
\label{newton}
\ee
\be 
\alpha_{o} = {(4\pi)^{2/3} { 1\over f_o V_{o} M_{11}^6}}=
(4\pi)^{2/3} {1\over f_o} {\left( {M_{GUT} \over  c M_{11}}
\right)^6}= \alpha_{GUT}
\label{alphao1}
\ee

and 

\be \alpha_{h} = {(4\pi)^{2/3}{1\over f_h} \left({  V_{h}^{-1/6} \over
M_{11}}\right)^6}
\label{alphah1}
\ee
Here $\langle V\rangle$ is the average volume of the Calabi-Yau space. In the linear 
approximation of (\ref{volo}) and (\ref{volh}), we have:
\be
\langle V\rangle=\frac {V_o +V_h}{2}
\label{averv}
\ee

The constant $f_o$ ($f_h$) is a ratio of  normalization of the traces of
adjoint representation of $G_o$ ($G_h$)  compare to $E_8$ case
\cite{witten85,nano1}. For a given compactification, $a_{o,h}$,$f_{o,h}$  
and $c$
are  taken to be fixed.  The theory has then  three free parameters $M_{11}$,
$V_6$ and $\rho$ which  in principle  could be determined as to fit
the experimental value of $G_N$ and  the
theoretical predictions for $M_{GUT}$ and $\alpha_{GUT}$. However
some values of these parameters might lead to a negative $V_{h6}$ or a very small value of $\pi \rho$ and
thus we will discard them as their physics is outside the domain of validity of our approximations.

 From equations (\ref{alphao1}) and (\ref{newton}), using $\alpha_{GUT}=1/25$,
and $M_{GUT}=2\cdot 10^{16}$GeV, one gets: 
\be 
M_{11}\sim 2.2 \frac {M_{GUT}} {f_o^{1/6} c}= \frac {4.5\cdot 10^{16}}
{f_o^{1/6} c}{\rm GeV}
\label{monze1}
\ee 
and 
\be 
\rho \langle V \rangle \sim (1.9 \  10^{16}\  {\rm GeV})^{-7} {(f_o^{1/6} c)}^9
\label{rove}
\ee 
As $c\le 1$ and $1.1\le f_o^{1/6}\le 1.35$ \footnote{ $f_0=$ 2.5,
3.75, 6 for $E_6$, $SO(10)$ and $SU(5)$ respectively.}, this relation
implies that $M_{11}$ is of the order of $M_{GUT}$ or a factor 2
larger.

In absence of fivebranes, the parmaters $a_o$ and $a_h$ are given by:

\be a_o =  \int_{CY} \omega \wedge \frac{ tr(F_o \wedge F_o) -
\frac{1}{2} tr(R \wedge R)}{8 \pi^2}
\label{ao}
\ee
and  
\be 
a_h = \int_{CY} \omega \wedge \frac{tr (F_h \wedge F_h) -
\frac{1}{2} tr (R \wedge R)}{8 \pi^2}
\label{ah}
\ee  where $\omega$ is a Kahler two-form of the Calabi-Yau
manifold\footnote{ The internal space is Kahler at the lowest order
in $\rho M_{11}^{-3}/V^{2/3}$ and  it is this lowest order volume
$V$ that appears in equations (1) and (2).}. To compute these
quantities we need to specify  the vacuum configurations of the gauge
fields on the two boundaries.  This is done by specifying the gauge
sheaves  $v_{o}$ and $v_{h}$ for the two $E_8$s.  

The four form  equations lead to  the topological constraint
that: 
\be 
 \frac {1}{8\pi^2}\left(tr F_o\wedge F_o + tr F_h 
\wedge F_h -tr R\wedge R 
\right) 
\label{Gcoho}
\ee 
must vanish  cohomologically. This means that the second Chern
numbers  associated with a four-cycle of the Calabi-Yau tangent bundle $TM$ 
and the two gauge sheaves  $v_{o}$  and $v_{h}$ must satisfy 
(the $c_1$s vanish): 
\be
c_{2}(TM)=c_{2}(v_o)+c_{2}(v_h) 
\label{ctwoeq}
\ee 
In the models considered below this implies that: 
\be 
a_h= -a_o 
\label{ahao}
\ee
In this case the average volume of the Calabi-Yau manifold appearing in 
(\ref{newton}) is identical to the lowest order value $\langle V \rangle=V$.

A detailed discussion of the different scales involves use of explicit values 
of the $a_{o,h}$ parameters \footnote{All previous numerical studies have
assumed $a_o\sim 1$ or $a_0=0$.}. As these are model dependent 
and need to be discussed.

\subsection{ Living on the edge with weakest gauge coupling constant}

Let's first discuss the situation where $a_o>0$. This is the case for example 
of Calabi-Yau's with standard\cite{CY1} and some the non-standard\cite{CY2} 
embeddings. In this case  $V_o >V_h$ and thus $\alpha_o < \alpha_h$: 
the standard model gauge bosons
and matter live on the wall of the universe with weakest coupling at
the unification scale. It was observed by Witten that as $V_o$ is
fixed, the relation  (\ref{diffvol}) implies that the coupling
constant become strong and the volume of the Calabi-Yau on the hidden
world becomes of the order of the Planck  length and low energy
supergravity approximation breaks down. Let's  estimate the different
values of the scales of the theory at this critical point. Taking $V_h= 0$
we find that $<v>=V=\frac {V_0}{2}$ and from (\ref{rove}) we get for the inverse size of the $S^1/Z_2$ segment:
\be
\frac{1} {\pi\rho_c}\sim \frac{M_{GUT}}{9.4 c^3 f^{3/2}}\sim \frac {2.1\cdot 10^{15}}{c^3 f^{3/2}}{\rm GeV}
\label{rocr}
\ee
For $G_o=E_6$ and taking $c=1$ we obtain $\frac {1}{\pi\rho_c} \sim 5.4\cdot 10^{14}{\rm GeV}$.  
Plugging the value (\ref{rocr}) in (\ref{volo}) we can obtain the necessary value for $a_o$ to solve our set of constraints:
\be
a_o\sim 1.3
\label{aorcr}
\ee
Away from the critical point, varying $V \ne \frac {V_0}{2}$, it is easy top see that it is possible to solve our constraint if:
\be
a_o\le a_{omax}\sim \frac {48}{\left(3125 \pi^4\right)^{1/3}}
\frac{M_{11}^3 V_o^{5/3}}{\rho V}\sim 1.7
\label{aomax}
\ee
These are in agreement with previous observation that for $a_o=1$ the desired value for $\rho$ is close to its maximum $\rho_c$.

The condition (\ref{aomax}) is very constraining for model builders. We were
 not able to find simple examples satisfying this constraint.

Some words of caution: 

(i) This solution with linear variation of the volume was derived assuming supersymmetry and lowest order factorization of the internal space as a segment 
times a compact internal . It should remain a good approximation as long as 
supersymmetry breaking scales (hidden and observable) are small compare to the size of the segment.

(ii) In the limit of $\rho \rightarrow \rho_{max}$ the
expansion parameter is large ${\rho  M_{11}^{-3}/V^{2/3}} \sim 1$ and our 
expansion (and thus the numerical values obtained from it) can not be trusted. However, Witten\cite{witten} has argued that there is a value of $\rho$ where 
the size of the hidden world becomes the order of the Planck length. 

(iii) The dynamics of the theory at $\rho \sim \rho_{max}$ is still unknown.

\subsection{Non-standard embedding: living on the edge with strongest coupling}

An issue that has not been discussed previously in the litterature is 
the possibility of 
having the ordinary particles living on the wall with weakest coupling.
Consider embedding the example of Calabi-Yau manifold of \cite{kach}
in the Ho\v{r}ava-Witten model. This space is defined as the intersection of vanishing loci of two  
degree six equations in $WCP^4_{1,1,2,2,3,3}$ space from which it inherits 
 Kahler form $J$ of the ambiant $CP^4$.  The gauge vacuum configuration is
specified by the gauge sheaves: 
$v_o=(2,2,2;1,1,1,1,1,1)$ and $v_{h} = (7;1,1,1,2,2)$ leading to an $E_6$
with three families in the observable sector and $SU(5)$ with 54 families (!) in the hidden sector. 
For this model:
\be
a_o= (c_2(v_o)-{1\over 2} c_2(TM))\int J\wedge J \wedge J 
= {1\over 2}(c_2(v_o)-c_2(v_h)))=-8\; ,
\ee
and
\be
a_h= (c_2(v_h)-{1\over 2} c_2(TM))\int J\wedge J \wedge J 
= -{1\over 2}(c_2(v_o)-c_2(v_h)))=8\; ,
\ee

The sign of $a_o$ and $a_h$ have been reversed compare to the previous example.
 The volume of the Calabi-Yau in the observable sector is now 
smaller that the one on the hidden wall. There is no maximal value of the 
length of the fifth dimension as we fix $V_o$ but let $G_N$ vary as was done 
in section 2.1. Some models might lead to very small $\rho$ and we might 
dismiss them.   
Solving our constraints leads to: 
\be
\frac {1}{\pi \rho} \sim\ 5.5\cdot 10^{15} {\rm GeV}
\label{param}
\ee

This toy example is not realistic as the the hidden sector coupling constant is small $\alpha_h\sim 0.001$ and the number of generations is huge. However we believe that this possibility merits more attention.

One can investigate the resulting scales for other examples of Calabi-Yau's 
with standard and  non-standard embedding as well as models with 
non-trivial $\pi_1$ or possibilities for $a_0=0$\footnote{ In this 
case $\frac {1}{\pi \rho} \sim  \frac {4.2 \cdot 10^{15}}{c^3 f^{3/2}} {\rm GeV}$.}. Results will be  
given elsewhere\cite{benf}.

\subsection{Exotics I: Five-brane impurities:}

The two coefficients $a_0$ and $a_h$ have opposite values. This is due to the 
constraint (\ref{Gcoho}). In the presence of fivebranes this constraint 
becomes\cite{witten,DMW} 
\be 
\frac {tr F_o\wedge F_o + tr F_h\wedge F_h -tr R\wedge R}{8\pi^2} +\Sigma_i [C_i]
\label{Gcoho2}
\ee 
must vanish cohomologically where where $[C_i]$ is the Poincar\'e dual cohomology of the holomorphic curve $C_i$ on which the fivebrane is located. If on the 
fifth direction the five-branes are located on the boundaries then the relation (\ref{ahao}) is still true. However if there an arbitrary set of fivebranes located in the bulk, the linear dependence in the bulk might change and the
 equations of motion should be solved for the new configuration.

\subsection{Exotics II: Gauge symmetries in the bulk}

While generically not present in the bulk, gauge symmetries might arise when 
the six dimensional internal manifold becomes singular if a small instanton 
is sitting on the singularity \cite{DMW,SW}. As discussed in\cite{sharpe} 
there are two
ways the $Z_2$ symmetry may act only in the fifth coordinate as it is the
case in the rest  of this paper. In that case the vector fields are odd. 
They are projected out and there is no new gauge symmetry at low energies. 
Neutral chiral 
multiplets are left in the boundaries and the gauge symmetry act on them 
as a global symmetry.

Another possibility is that the $Z_2$ projection is supplied with a freely 
acting $Z_2$ on the $CY$ space one might be able to preserve the gauge 
symmetry in the bulk. There are three phenomenological scenarios that one
might consider: (a) The ordinary matter lives on one boundary. 
(b) The ordinary matter lives in the bulk. Supersymmetry might be broken
by non-perturbative effects on the boundaries. The two potentiels 
might stabilize the moduli because of the difference dependences in the 
potentials. (c) Part of the ordinary matter is on the boundaries and part in the bulk.

One may also speculate about the possibility that the non-perturbative and the 
perturbative gauge groups mix. In that case after $Z_2$ projecting out the gauge bosons  one may 
generates a ``global'' symmetry acting on the ordinary matter that might for 
example be used to solve the problem of fast proton decay. In the case (a)
the gauge bosons in the bulk might play the role 
of mediators of supersymmetry breaking at low energies.  
 While it is useful to keep in mind these possibilities, in absence of 
concrete models 
these remain pure speculations and we do not discuss further these issues.

\section{Life on the edges and in the bulk of the M-world}

\subsection{Dark matter}

Many astronomical observations seem to to indicate that most of the mass 
of the universe is  invisible\footnote{In the sense of not 
emitting photons but of course it is visible through gravitational effects.}. 
While part of it might be simply vacuum energy (cosmological 
constant), a big mass (at least 40 times the one of the observed part) seems 
to be made of some ``dark'' matter. Here we would like to 
discuss the possibility of such matter arises from the hidden wall or 
the bulk of M-theory on $S^1/Z_2$\cite{BEN}.

The early expansion of universe might have been different on the two
boundaries \cite{kar1}. This anisotropy however disappears as it gets
washed away by gravitational forces at late times\footnote{ There is 
the possibility that part of hidden matter masses keep quickly varying with 
time leading to anisotropies.}.  
We suppose that after a  period of inflation we are left with an
isotropic universe with a single Hubble ``constant'' $H_0$ and a standard
thermal history of the universe. The inflaton will decay into the
ordinary particles  and the hidden ones with a widths $\Gamma_o$ and
$\Gamma_h$ respectively reheating the boundaries to temperatures
$T_o \sim \Gamma_h^{1/2} G_N^{-1/4}$ and  $T_h\sim \Gamma_h^{1/2} G_N^{-1/4}$. 
The gravitational interactions are weak and do not thermalize the two baths. 
If $\Gamma_h \sim 0$ the hidden universe is cold, empty and (except for
supersymmetry breaking) has no influence on the  large scale structure
of the universe. 

As pointed out in \cite{KT}  agreement between the mass of earth
computed from seismic data and the one computed from kinematics of 
its satellites on one hand and 
on the other hand difficulties to modify solar models, imply that the density 
of  these particles is small in the close neighborhood of earth. However 
this does not exclude the possibility that hidden worlds provide a large fraction of the  dark matter in the universe.

Let's discuss (in a qualitative way) the properties of the particles 
generically present in the model and the possibility that they might provide 
us with dark matter candidates.

\subsection{Particles living on the hidden wall:}

{ \bf Light} hidden particles with masses less than ${\cal O}$(MeV) would  
contribute to the effective degrees of freedom $g_{eff}$ that determines 
the energy density (thus the expansion rate) of the universe during the 
primordial nucleosynthesis period\cite{KT}:
\be
g_{eff} = g_* ( 1 +  \frac {g_{*h}}{g_*}\left( \frac {T_h}{T_o}\right)^4 )|_{t=t_N}
\label{geff}
\ee
where $g_*$ and $g_{*h}$  are the degrees of freedom of ordinary 
particles and hidden ones, $g_*=10.75$. All the quantities are computed 
at the Big-Bang 
nucleosynthesis time  $t_N$. Using  $g_{eff} \le 13$, one could arrive at
a constraint on $T_h$ for specific models where $g_{*h}$ is known or on  
$g_{*h}$ if $T_h$ is known. This constraint leads in general to $T_h < T_o$. 
The possibility of such light particles being a large part of dark matter 
depends on their abundances which is very model dependent. 

{\bf Heavy} hidden particles might arise as bound states of a non-abelian group
confining in the hidden wall\cite{BEN}. Their mass is expected to be roughly 
of the order of the scale of confinement:
\be
\Lambda \sim V_h^{-1/6} e^{-b/\alpha_h}
\label{lam}
\ee 
where $\alpha_h$ is given in (\ref{alphah1}).

Let's suppose that in analogy with the case of proton in the observable world
who are stable (or to the ``cryptons'' of weakly heterotic strings 
that have been claimed to have
a lifetime bigger than the age of the universe\cite{cry}), the lightest of 
the bound states $X$ is cosmologically stable. 
In \cite{BEN} it was suggested that  discrete symmetries would 
be responsible of this stability in a similar way to the case of the 
proton\cite{BD}.

If the confining gauge group is the same responsible for supersymmetry 
breaking then  the mass of $X$, $m_X$ is roughly of the order of 
\be
m_x \sim \Lambda \sim 10^{11 -14}{\rm GeV}  
\label{m_x}
\ee
 
Consider the possibility that there is  such state $X$  weakly 
interacting and $m_x \sim H_0$. Weakly stands here for $n_{X} \langle \sigma_A |v| 
\rangle \le H$, where $n_X$ is the (conserved) co-moving number density, $H$ is the Hubble constant and
$|v|$ is the Moller speed for the dark matter particles.  Notice that the only constraint that 
might be hard to satisfy is stability and it is model dependent. Under these 
conditions $X$ might have never been and will never go in local thermal 
equilibrium. 

The computation of the abundance of such kind of particles have been recently
reconsidered \cite{KRT,CKR,kuz1}. The numerical computations of 
\cite{CKR,kuz1} seem to indicate that interactions of the vacuum with 
the gravitational field might lead to the creation of the desired amount of 
$X$ particles so that it constitute a large fraction of the mass of the 
universe.

The $X$ particles do not decay into ordinary matter and thus
can only be detected through their gravitational effects. Their experimental 
signature would be abundance of black holes and dark massive objects that 
might be detected in experiment based on gravitational lensing. We will return to these issues in\cite{benf}.

\subsection{Particles living in the bulk}

 In addition to the massless supergravity multiplet one has massive modes. 
In the models where the fifth dimension is much larger than the other six compactified ones, the lightest massive modes,  the Kaluza-Klein (KK) states associated with the fifth dimension, have masses  given by:

\be
m_n\sim \frac {n}{\rho} \sim n \cdot 10^{15-16} {\rm GeV}
\label{mKK}
\ee

Let'us first discuss the quantum numbers of these particles. Consider the $N=2$
theory obtained through compactification of eleven dimensional supergravity on $CY \times S^1/Z_2$. Under the action of $Z_2$ the fields are divided into 
even ones 
\be
\psi_{even} \rightarrow  \psi_{even}
\label{psie}
\ee
which form multiples of an $N=1$ supergravity four dimensional theory and odd ones
\be
\psi_{odd} \rightarrow -\psi_{odd}
\label{psio}
\ee
for the remaining of the fields. In particular all the vector fields on the 
bulk are projected out as they originate  in eleven dimensions from
 the three form  which is odd under $Z_2$.

Now at the level of massive states all the quantum numbers survive\footnote{Physics of KK states in orbifolds has been investigated in\cite{kk}.}. This 
because translation invariance is broken in the fifth direction and propagating states do not have a definite momentum but are of the form:

\be
\psi_{even} \times \left( |\frac {n}{\rho}> + |\frac {n}{\rho}>\right)
\; \; \; \; {\rm and} \; \; \; \; \psi_{odd}\times\left( |\frac {n}{\rho}> - |\frac {n}{\rho}>\right)
\label{states}
\ee

We do not expect the bulk states to carry new quantum numbers that are 
not present in  the twisted sector. KK states are instable and they decay 
with a lifetime: 
\be
\tau_{KK} \sim \frac{\rho^3}{G_N}
\ee
for those which decay through gravitational strength interactions.
One expect a correction to the strength of the coupling constant roughly going as $e^{-n/\rho^2}$ due from the Euclidean action necessary to extract momenta 
from the vacuum at the fixed points. However such corrections are small 
for large $\rho$. For $\rho \sim 10^{15}$ GeV we find $\tau_{KK}\sim 10^{-40}$
second. In the simplest class\footnote{ This is not the case if 
non perturbative gauge symmetries arise live in the bulk(see section 2.4).} of compactifications of Ho\v{r}ava-Witten theory the bulk does not provide natural candidates for dark matter.

Finally I will end this section by a comment on  extreme cases. If the 
seven dimensional internal space is very deformed with  the Calabi-Yau 
charateristic length becoming large at some point of the segment or 
anisotropic, the massive modes are described by localized states 
as no translation invariance remains along the fifth direction.

\subsection{View from Dr. Jekyll's and Mr. Hyde's powerful microscopes}

We consider the geometry of space-time when we increase the probe energy 
as seen by Dr. Jekyll living on the observable world and Mr. Hyde living in 
the hidden world. 

At low energies the world is four dimensional. Both Dr.Jekyll and Mr. Hyde 
are not 
aware of existence of higher dimensions. All observations in nature are 
described with four 
dimensional  representations and parameters. But the latter do not seem 
neither simple nor natural. Interactions between the two worlds are 
with gravitational strength. Only at the cosmological scales, where all the 
other interactions are screened, that the two universes influence each other.

With the increase the probe energy and luminosity of his particle colliders 
Dr. Jeckyll 
studies some very energetic process where (thanks to the very good precision  
of his instruments) he begins to observe deviations from the predictions of 
the minimal supersymmetric model.
To explain the observations he needs to take into account 
gravitational forces. If the precision is better than $10^{-14}$, this
happens at energies around $10^{14}$ GeV. 

Time to time, if in the hidden  world there are particles with lifetime long 
enough to escape , he observes some 
missing energy events. He has created particles on the hidden wall. Through his ``gravitational detectors'' he can observe the new particles. 

At energies of the order of $10^{15}$ GeV Dr. Jekyll observes that the 
strength of 
the gravitational interactions grows faster. The rate of production 
of hidden particles increases. He realizes that the world is five-dimensional.
This puts some order in the parameters of his models: they look more natural.

At energies of the order $10^{16}$ GeV all the interactions have the same 
strength. New massive modes begin to contribute to the interactions, the 
world looks eleven dimensional...If the precision of Dr Jekyll's apparatus 
is worse than $\sim 10^{-4}$, it is only very close to $10^{16}$ GeV energy 
scales that he finds out that he needs to include gravitational effects and 
the these should be in a five-dimensional space-time.

Depending of the strength of the interactions weaker or stronger in the hidden world, Mr. Hyde (if stable) might have been aware 
of the eleven dimensions before or after Jekyll respectively.

\section{Cosmological solutions:}

\subsection{Generating cosmological solutions from p-branes:}

A simple way to generate  cosmological solutions is to start with
p-brane solutions in D-dimensions. These lead to a space with a 
metric of the form:
\be
ds_D^2= ds_{p+1}^2+ ds_{D-p-1}^2
\label{met1}
\ee

The $ds_{p+1}^2$ is the world-volume part: the space-time swept by the brane 
when it propagates. It contains a time coordinate. Our first step is to 
Euclideanize this coordinate and may be (if we like so) compactify this space.
 We have then an Euclidean solution in $D$-dimensions. Next we remark that 
the metric
contains factors with explicit dependence on the coordinate of the transverse 
space. If the p-brane solution has a ``spherical" symmetry in a part of 
the transverse 
space then the dependence of the metric will be on the radial coordinate 
associate with this space. Thus our second space is to rotate this radial 
coordinate into a time-like direction. We have thus generated a solution 
with time dependent metric. To get a time which varies from $-\infty$ to 
$+\infty$ one needs to make a simple changement of variables. As p-branes
metrics interpolate between the horizon geometry and some space-time geometry 
at infinity, our process generates a cosmological solution that interpolates 
in time between the two geometries\cite{interp}. 
 We would like to discuss briefly some examples of such solutions in M-theory.
A long list of references on cosmological solutions can be found 
in \cite{kar1}.

\subsection{ An example in M-theory on $S^1/Z_2$ }

One starts with a five-brane solution of the Ho\v{r}ava-Witten $M$-theory\cite{witten,ovr,dud}. The metric on this space is of the form:

\beba
ds^2 &=& e^{2A}\left( -dt^2+ dx^\m dx^\n\d_{\m\n}\right) + \nn\\ &&
e^{2B}\left( d r^2 + r^2 \left( d\phi ^2 +\sin^2\! \left( \phi \right)
\left( d\theta^2 + \sin^2 \!\theta d\omega^2\right) \right) \right)
\nn\\ &&+  e^{2B} (dx^{11})^2,
\label{oldm}
\eaee
where the functions $A$, $B$ are  related by $A=-B/2=-C/6$ and the four-form 
$G$ is given by:  
\be 
G_{mnrs} =
\pm\frac{1}{\sqrt{2}}e^{-\frac{8C}{3}}{\epsilon_{mnrs}}^t\partial_t
e^C\;
\label{ffans}
\ee 
with $m,n,...=6,...,9,11$. The function $C$
depends only on  $x^{11}$ and $r$. On the boundaries the four-form gets 
contributions from the gauge vacuum configuration. More precisely 
\footnote{ $\k^2 =M_{11}^9$ and $\frac {\l^6}{\k^4}=\left( 4\pi \right)^5$
\cite{witten,LU}}:
\bea 
\left. G_{ABCD}\right|_{x^{11}=0} =  -\frac{3}{\sqrt{2}}
 \frac{\k^2}{\l^2} \mbox{tr}(F^{(1)}_{[AB}F^{(1)}_{CD]})\nn\\
 \left. G_{ABCD}\right|_{x^{11}=\pi\r} =  +\frac{3}{\sqrt{2}}
 \frac{\k^2}{\l^2} \mbox{tr}(F^{(2)}_{[AB}F^{(2)}_{CD]})\; .
 \label{Bianchi3} 
\eea
Notice that the solution is independent on the time $t$ coordinate and the 
latter can be taken to be compact. More over a rotation to 
Euclidean space leads to a theory with real parameters in the action. 
By rotating the radial
coordinate into a time like direction we can achieve the  
changement of variables:
\be 
t=\sqrt{q}x^0 \, \ \ \ \ \, \phi= \sqrt{q} \chi \ \ \ \ \,
r=\frac{\t}{ \sqrt{q}}. 
\label{varc}
\ee 
on the original solution, where $q$ is taken from $+1$ to $-1$.

This leads to a cosmological solution\footnote{This solution corresponds to 
a   strong heterotic string coupling  limit  of a solution considered 
 in \cite{bfs}.} of the form\cite{kar1}:

\begin{equation}
ds^2 = \Phi^{-1/3}\left( (dx^0)^2+dx^\m dx^\n\d_{\m\n}\right) +
\Phi^{2/3}\left( d\t^2 + \t^2 d\Omega^2 _{3,-1} \right) +  
\Phi^{2/3} (dx^{11})^2,
\label{finmet}
\end{equation}
with 
\begin{equation}
d\Omega_{3,k}^2 = d\chi ^2 +\frac{\sin^2\! \left( \sqrt{k}
\chi\right)}{k} \left( d\theta ^2 + \sin^2 \!\theta d\omega^2\right).
\end{equation}
In (\ref{finmet}) $\Phi$ is given by: 
\be \Phi = \Phi_0+\phi, 
\label{ansa} 
\ee 
where $\Phi_0$ 
is a function of $\t$ only:

\be 
\Phi_o = 1 + \frac{2 \k^2}{\pi\r \lambda^2} \left ( \frac{2
\sigma_1^2+ \t^2}{(\sigma_1^2 + \t^2)^2} + \frac{2 \sigma_2^2+
\t^2}{(\sigma_2^2 + \t^2)^2} \right ) \; . 
\label{fio}
\ee 
and $\phi$ is the perturbation dependent on $x^{11}$. It can be written as\cite{ovr}:

\be
\phi = \sum_{n=0}^{n=\infty} \phi_n\label{phi}
\ee
 The leading order of the expansion is given by:

\be \phi_0 = 48 \frac{\k^2}{\lambda^2} \left ( P_0 (x^{11})
\frac{\sigma_1^4} {(\sigma_1^2 + \t^2)^4}+  Q_0 (x^{11})
\frac{\sigma_2^4} {(\sigma_2^2 + \t^2)^4} \right )  
\label{smphio}
\ee 

where:
\be P_0 = - \frac{(x^{11})^2}{4 \pi\r} + \frac{x^{11}}{2} - \frac{\pi
 \r}{6}\; , \quad Q_0 = - \frac{(x^{11})^2}{4 \pi\r} + \frac{\pi
 \r}{12} 
\label{eq:pzero}
\ee 

To have a time
coordinate which goes from $-\infty$  to $+\infty$ one makes the
rescaling  $\tau=e^{\eta}$. The choice $\tau=e^{-\eta}$ lead to a
similar branch  disconnected from this one. A four
dimensional solution is obtained by considering the six-dimensional transverse
space parametrized by $x^0,...,x^5$ as compact.

For $\t\to 0$ or equivalently $\eta\to -\infty$ the size of the compact
internal space is finite. However the radius of the pseudo-sphere of
the universe goes to zero. Through tunneling effect a bubble with a
new expectation value of the field $\Phi$  (the  dual of the four-form
$G$) is created at $\eta\to -\infty$ and expands. 
This event is at finite proper time 
from any point in
the future. If one start with different  initial conditions on the
 boundaries  (  $\s_1 \neq \s_2$), the expansion on the two
boundaries follows different time dependence. On the $x^{11}=0$
boundary :

\bea 
\Phi &=& 1 + \frac{2 \k^2}{\pi\r \lambda^2} \left ( \frac{2
\sigma_1^2+ \t^2}{(\sigma_1^2 + \t^2)^2}+ \frac{2 \sigma_2^2+
\t^2}{(\sigma_2^2 + \t^2)^2} \right )\nn \\ && +  8 \frac{\k^2\pi \r}
{\lambda^2} 
\left ( 
- \frac{\sigma_1^4} {(\sigma_1^2 + \t^2)^4}+  
\frac{\sigma_2^4} {2(\sigma_2^2 + \t^2)^4} \right )+... \; ,
\label{edge1}
 \eea  

while for the hidden universe $x^{11}=\pi \r$:
\bea
\Phi&=& 1 + \frac{2 \k^2}{\pi\r \lambda^2} \left ( \frac{2
\sigma_1^2+ \t^2}{(\sigma_1^2 + \t^2)^2} + \frac{2 \sigma_2^2+
\t^2}{(\sigma_2^2 + \t^2)^2} \right ) \nn \\ && +  8 \frac{\k^2\pi \r}{\lambda^2} 
\left ( 
\frac{\sigma_1^4} {2 (\sigma_1^2 + \t^2)^4}-  
\frac{\sigma_2^4} {(\sigma_2^2 + \t^2)^4} \right )+... \; ,
\label{edge2}
 \eea  
where the dots represent higher orders of the expansion in $(\pi\r)^2/\s^2$.

At later times as $\t\to +\infty$ or equivalently $\eta\to +\infty$, we
find $\Phi \to 1$. Thus the solution describes an expanding four
dimensional flat universe. The expansion anisotropy between the two
boundaries is washed away by gravitational interactions between
them. It is also interesting to notice that the volume of the internal
space and the size of the segment remain finite (small) and thus we
have a mechanism of spontaneous compactification.

 Our field $\Phi$
(obtained through dualization of the four form) or more precisely
$\Phi_0$ would the analogous of the scalar field considered in \cite{haw} 

We have  illustrated the possibility of having different initial
 conditions on the two boundaries of the universe which might lead to
 different expansions at early time.

\subsection { Other examples of nucleation of universe in M-theory}

In the previous example  the  asymptotic geometry as $\tau \rightarrow
\infty$ is a flat space-time. Nucleations of universes that are described 
through rotating instanton
solutions in Euclidean space-time into a tunneling effect in spaces with 
Lorentzian signature
have attracted a lot of interest recently\cite{haw}. 

I would like to comment on the possibility to obtain solution which at 
late time become asymptotically an
anti-de-Sitter space. Details will be provided in\cite{K2}.

To simplify the discussion, consider  an extremal p-brane solution in a $D$-dimensional  Minkowski space. The metric takes the form:

\be 
ds^2= H^\alpha d{x_{||}}^\mu d{x_{||}}_\mu +  
H^\beta d{x_{T}}^a d{x_{T}}^a 
\label{harm}
\ee
where $x_{||}$ and ${x_{T}}$ are the coordinates of the parallel and 
transverse spaces to the world-volume of
the p-brane. The harmonic function $H$ depends on the radial 
coordinate of the space transverse to the p-brane. Generically
\be
H\sim h+{c \over r^{\tilde d}}
\label{harm2}
\ee
where ${\tilde d}=D-p-3$ , $c$ is related to the charge carried by the 
p-brane and $h$ is a constant of integration taken to be $h=1$ so that 
the asymptotical  geometry  as $r\rightarrow \infty$ is a flat 
Minkowski space-time.

Recently it has been argued\cite{hyun} that there is a set of 
transformations
that shifts the value of $h$. Starting with $h=1$ these transformations
lead to the metric \ref{harm} with the difference that 
now $h=0$. This allows us to get solutions 
which asymptotically as $r \rightarrow \infty$ go to an anti-de-Sitter geometry. After exchange of the role of 
time and transverse radial coordinate, we obtain\cite{K2} cosmological 
solutions expanding toward  anti-de-Sitter space. Some of these
 solutions have instanton interpretation in a similar way to the example of the 
previous subsection\cite{K2}.

One could  have instead started with a solution on a
curved background. For instance one could transform the membrane solution
of\cite{mbrnew} and generate a new cosmological solution: at time goes to infinity the space-time geometry goes to four-dimensional anti-de-Sitter space with the membrane at the end of the universe\cite{K2}.

\section{Conclusions}

We have described three issues related to cosmological applications of 
M-theory. We have investigated in great details the question of scales 
as it is necessary prior to
any discussion of physical phenomena. The geometry of the 
space-time is a product of a flat four dimensional Minkowski 
space-time and a seven 
dimensional compact space. The latter is approximated as a fibration of 
a Calabi-Yau space on a 
segment. The volume of the Calabi-Yau varies slowly (as to satisfy  the adiabatic principle\cite{vw}) and the theory  describes strongly coupled limit of 
heterotic 
string theory. A linear approximation was  used to extract orders of 
magnitude of the characteristic sizes for the Calabi-Yau space and the fifth 
dimension.  We investigated the possibilities that three possibilities 
that $\alpha_o< \alpha_h$, $\alpha_o> \alpha_h$ and $\alpha_o = \alpha_h$.
Only in the first class we found that (if the value of the 
Calabi-Yau volume on the observable 
boundary is fixed) the length of the fifth dimension has a maximal value.

We  discussed the possibility that dark matter originates from the bulk or  on the hidden wall of the theory. We found that the former is unlikely 
while the latter is an open possibility that needs to be investigated in more details\cite{benf}. We discussed some model-independent properties of this 
particles.

Finally we presented a toy model for a cosmological solution describing an instanton effect in M-theory on $S^1/Z_2$ framework. This has the important 
features that it exhibits an example for a scenario where internal dimensions 
remain small 
while the four dimensional space expands. Moreover the expansion is anisotropic in the fifth direction, a situation we believe is generic to 
Ho\v{r}ava-Witten class of models.

\section*{Acknowledgments}
I wish to thank  M. Benakli, P. Candelas, M.J. Duff, J. Ellis, I. Lavrinenko, J.-X. Lu, D.V. Nanopoulos, C.N. Pope, E. Sharpe and  T.A. Tuan for useful discussions.  This work was supported  by DOE grant 
DE-FG03-95-ER-40917.


\end{document}